\documentclass[twocolumn,showpacs,aps,prl]{revtex4-1}

\usepackage{epsfig}
\usepackage{latexsym}

\begin{document}
\title{Wilson's numerical renormalization group and AdS$_3$ geometry}
\author{Kouichi Okunishi}
\affiliation{Department of Physics, Faculty of Science, Niigata University, Niigata 950
-2181, Japan.}

\date{\today}
\begin{abstract}
We discuss the relation between the Wilson's numerical renormalization group(NRG) for the Kondo impurity problem and a field theory in the background AdS$_3$ space time,  where the radial coordinate plays a role of the controlling  parameter of the effective mass scale. 
We find that the Wilson NRG can be described by the boundary Rindler field and then the cutoff parameter $\lambda$ of the Wilson NRG is related to the AdS radius $L$ through $\lambda = 2k_F/\omega L$, where $k_F$ is the effective Fermi wave number.
It is also found that the Rindler space is discretized with the lattice space of $a=\pi/k_F$.
\end{abstract}


\maketitle

Recently,  geometrical approach for quantum many-body systems attracts considerable interest in accordance with the development of AdS/CFT\cite{adscftreview}.
In particular, the area law of quantum entanglement has been strikingly demonstrated by the concept of holography through extended space-time\cite{ee}.
On the other hand, condensed-matter theory often provides an interesting play ground of the concept that may be  difficult to directly observe in high energy physics.
 Applications of AdS/CFT have been actually performed to extract the low-energy behavior of a certain class of condensed matter physics such as holographic superconductivity\cite{holographicsuper}.
More recently, it is pointed out  that multiscale entanglement renormalization ansatz(MERA)\cite{MERA}, which is an efficient numerical renormalization group(NRG), involves a geometrical structure satisfying the area low of the entanglement entropy\cite{MERAEE}.
The connection between the holography and renormalization group(RG) has been an essential subject to understand physics behind the quantum many body system also form the condensed matter view point.

Among various RGs in condensed matter theory,  Wilson's approach for the Kondo impurity problem\cite{Kondo} is a milestone of NRG, which enables for us to quantitatively extract the low-energy physics in the metallic system\cite{Wilson,bulla}.
In the Wilson NRG, the low-energy excitation around the Fermi surface is logarithmically discretized for numerical computation and  the resulting effective Hamiltonian is reduced to be the 1+1 dimensional(D) tight-binding electrons with the exponential deformation of the electron hopping amplitude.
An interesting point is that this effective Hamiltonian has the scale free property associated with the lattice translation, which is a key point of success of the Wilson theory\cite{ON}.
However, the physical meaning of the discretization cutoff parameter $\Lambda$ that is introduced by numerical reason is unknown, although the energy scale controlling of the Wilson NRG significantly relies on this cutoff parameter.

The purpose of this paper is to reveal a connection of the Wilson's effective Hamiltonian to the background AdS$_3$ space-time.
We will point out that the Wilson NRG can be related to the Rindler filed\cite{fulling, unruh, rindler}  in the level of the equation of motion.
We further introduce the  AdS$_3$ space time, where the radial coordinate plays a role of controlling the energy scale of the boundary 1+1D system.
We then find that the AdS radius $L$ can be related to the deformation parameter of the Wilson NRG, and the boundary Rindler space should be automatically discretized with a lattice space associated with the length scale in the UV limit.

The Wilson's effective Hamiltonian with the exponential deformation is written as 
\begin{equation}
{\cal H} = \sum_n e^{-\lambda n} (a_n^\dagger  a_{n+1} + a_{n+1}^\dagger a_n)
\label{Hamiltonian}
\end{equation}
where $e^{\lambda} \equiv \Lambda$ is the cutoff parameter due to the logarithmic discretization\cite{sign}. 
In the Kondo problem,  $a_n$ is the fermion annihilation operator on the 1D chain.
In the following, however, we basically consider the single particle state, where the statistics of the particle is less important. 
Thus we deal with the scalar field for simplicity, and accordingly consider $c_n$ as a (hard-core) bosonic particle operator below.

Writing the single particle state as $|\psi \rangle =\sum_n \psi(n)a_n^\dagger |0\rangle$
and introducing $ \psi(n)= e^{\lambda n /2} \phi(n)$, we have the single particle Schr\"odinger equation 
\begin{eqnarray}
  \phi(n-1)+ \phi(n+1)= Ee^{-\lambda/2} e^{\lambda n} \phi(n), \label{eigenvalueeq2}
\end{eqnarray}
where $E$ is the energy eigenvalue\cite{ON}.
We take naive a continuum limit of Eq. (\ref{eigenvalueeq2}), writing $\xi=a n$ with the lattice space $a$. 
Introducing renormalized parameters $k_F$, $L=2 a/\lambda$ and ${\cal E}=2\sqrt{Ee^{^{-\lambda/2}}}$, we obtain the equation for a positive energy $E$ as
\begin{eqnarray}
&   \phi''(\xi) + ( k_F^2- {\cal E}^2 e^{2\xi/L}) \phi(\xi) =0  \label{eigenvalueeq1} .
\end{eqnarray}
Note that $\xi\to -\infty$ corresponds to the impurity site.
A significant aspect on this equation is that it is invariant under the extended translational symmetry, $\xi \to \xi+ c $ combined with ${\cal E}\to {\cal E}e^{c/L}$, which implies that the energy scale of the system can be controlled by the translation operation of $\xi$. 
Then, the inner product of the wavefunctions is defined as 
\begin{equation}
(\phi, \phi')\equiv \int_{-\infty}^{\infty} \frac{d\xi}{L} e^{2\xi/L} \phi^\dagger(\xi) \phi'(\xi),
\label{norm}
\end{equation}
with which the fields $\{\phi\}$ conform the orthonormal basis of (\ref{Hamiltonian}).
According to numerical calculation for the lattice Hamiltonian, the solution of (\ref{eigenvalueeq1}) well reproduces the eigenstate of the lattice model (\ref{eigenvalueeq2}) with fixing $ k_F\equiv  \pi/2a$. 
Thus we can assume that Eq. (\ref{eigenvalueeq1}) describes the correct orthonormal basis in the Wilson NRG problem.

We next consider a scalar field in the Rindler space time\cite{fulling}.
The relation between the Rindler coordinate ($\eta, v)$ and the 1+1D Minkowski space $(t, x)$ is given by
\begin{equation}
t=Le^{v/L}\sinh(\eta/L)\qquad x=Le^{v/L}\cosh(\eta/L)  
\label{rindlercoordinate}
\end{equation}
which represents physics observed from a constantly accelerating observer.
In the Rindler wedge of $x> |t| $, the Klein-Gordon equation is obtained as
$
 (\partial_\eta^2 -\partial_v^2  +m ^2e^{2v/L} )\phi =0
$.
Substituting mode expansion: $\phi(\eta,v)= e^{i\omega \eta} \phi(v)$, we have
\begin{equation}
( \partial_v^2 + \omega^2  -m ^2 e^{2v/L} )\phi =0 \label{rindlerkg1}
\end{equation}
What we should remark is that Eqs. (\ref{eigenvalueeq1}) and (\ref{rindlerkg1}) have the same form, which suggests that the Wilson NRG and the Rindler field partly shares the common background of physics. 
However, there is also an essential difference;
The usual Klein-Gordon theory determines the energy mode frequency $\omega$ for a fixed mass $m$.
 In the Wilson theory, on the other hand, the mass term becomes the energy eigenvalue ${\cal E}^2$, while the frequency term is fixed at the Fermi momentum $k_F$.
This suggests that the Wilson theory is interpreted as  a kind of the inverse problem of the usual field theory, i.e. a problem of determining adoptive mass scale ${\cal E}$ for a fixed frequency $k_F$.

Since the original free electron problem represents the massless excitation around the Fermi surface, it may sound slightly strange that the Wilson theory is related to a field theory containing the ``mass term'' ${\cal E}^2$. 
However, the scale of the ${\cal E}^2$ can change arbitrary for a fixed $k_F$  by the lattice translation\cite{ON}.
This suggests that the mass scale in the Wilson theory runs from the UV to IR scale along a certain trajectory connecting  a set of the Klein-Gordon fields in the mass parameter space.
In this sense, the Wilson theory may be regarded as a renormalization group for the mass parameter in an expanded space-time.

%---------------------------
%\section{extended space}

In order to formulate the above observation, let us consider the AdS$_3$ space as a background metric, which is often written by the  Poincar\'e coordinate
$
ds^2= \frac{L^2}{z^2}( dz^2 -dt^2 +dx^2 )
$.
We rewrite this coordinate through (\ref{rindlercoordinate}) and 
$
z=Le^{u/L}
$;
\begin{equation}
ds^2= du^2 - e^{2(v-u)/L}(d\eta^2 - dv^2 ) , \label{adsrindler}
\end{equation}
The equation of motion for a scalar field with this metric is given by
\begin{equation}
 (\partial_u^2 + \frac{2}{L}\partial_u-m^2)\Psi - e^{2(u-v)/L}(\partial_\eta^2 - \partial_v^2)\Psi =0.
\end{equation}
Decoupling the field into the radial and boundary modes, $\Psi=\Phi(u) \varphi(\eta,v) $, and performing the mode expansion $\varphi = e^{i\omega \eta} \phi(v)$, we obtain
\begin{eqnarray}
 \partial_v^2 \phi(v) + (\omega^2 - \mu^2 e^{2v/L})\phi(v) =0\label{radial0}\\
(\partial_u^2 + \frac{2}{L}\partial_u-m^2+\mu^2 e^{2u/L})\Phi(u)=0\label{rindler0} 
\end{eqnarray}
where the sign in front of $\mu^2$ corresponds to the positive energy solution.
In these equations,  the bare mass term $m$ appears only in the radial mode, while $\varphi$ represents the boundary Rindler field with an effective mass scale $\mu$.
Now, we rescale Eqs. (\ref{radial0}) and (\ref{rindler0}) through $v=\frac{k_F}{\omega}\tilde{v}$, $u=\frac{k_F}{\omega}\tilde{u}$, $\tilde{\mu}=\frac{k_F}{\omega}\mu$ and $\tilde{L}=\frac{\omega}{k_F} L$.
Then a couple of equations becomes
\begin{eqnarray}
 \partial_{\tilde{v}}^2 \phi + (k_F^2 - \tilde{\mu}^2 e^{2\tilde{v}/\tilde{L}})\phi =0 \label{rindler1}\\
(\partial_{\tilde{u}}^2 + \frac{2}{\tilde{L}}\partial_{\tilde{u}}-m_r^2+ \tilde{\mu}^2e^{2\tilde{u}/\tilde{L}})\Phi(\tilde{u})=0 \label{radial1}
\end{eqnarray}
where we define the renormalized mass 
\begin{equation}
m_{\rm r}\equiv \frac{k_Fm}{\omega}.
\end{equation}
For the usual 1+1D  Rindler field (\ref{rindlerkg1}),  $m^2$ is fixed at a certain mass and then $\omega$ conforms the single particle spectrum.
In contrast, a couple of equations (\ref{rindler1}) and (\ref{radial1}) has two quantum number for $\tilde{\mu}$ and $m_{\rm r}$, which are rescaled from $\mu$ and $\omega$.
Then, a particular point is that $k_F$ in Eq. (\ref{rindler1}) is fixed at the Fermi wave number and, instead, the ``mass term'' $\tilde{\mu}^2$ acquires spectrum for this fixed $k_F$.
In this sense, Eq. (\ref{rindler1}) can describe a desired situation of the inverse problem of the Rindler field theory.
 Moreover, we can see that the deformation parameter in the Wilson NRG is given by $\lambda = 2/\tilde{L}=2k_F/(\omega L)$.

Let us first discuss the solution of Eq. (\ref{rindler1}), which is explicitly given by the modified Bessel function of imaginary order,
\begin{eqnarray}
\phi(\tilde{v}) \sim K_{i k_F \tilde{L}} (\tilde{\mu}\tilde{L} e^{\tilde{v}/\tilde{L}}).
\label{solution1}
\end{eqnarray}
This solution behaves like a free particle $\phi(\tilde{v}) \sim e^{\pm ik_F \tilde{v}}$ in the $\tilde{v}\to -\infty$ limit.
In the Minkowski coordinate,  Eq. (\ref{solution1}) covers the half-infinite spatial region($x>|t|$) and then the boundary of $\tilde{v}=-\infty$, which is nothing but the impurity site, reduces to  $t=\pm x$.
Here recall that the impurity site corresponds to the UV limit in the Wilson NRG.
In term of the Rindler theory,  $\tilde{v}\to -\infty$ corresponds to the horizon for the uniformly accelerating observer.
Thus, the Kondo problem can be interpreted as the impurity spin interacting with the asymptotically free particle of $e^{\pm ik_F v}= (x^2-t^2)^{\pm i\omega /2}$ at the Rindler horizon\cite{fulling,unruh,rindler}.
The boundary condition of $\phi(\tilde{v})$ at the UV cutoff scale($\tilde{v}=\log \epsilon$ with $\epsilon \sim +0$) is important.
On the other hand, in the  $\tilde{v} \to \infty $ limit,  $\phi(\tilde{v})$ decays very rapidly, where the $\tilde{\mu}^2e^{2\tilde{v}}$ term in Eq. (\ref{rindler1}) becomes dominant.
Thus we have no difficulty for $\phi(\tilde{v})$ in the IR limit. 

For the fixed $k_F$,  Eq. (\ref{rindler1}) determine the eigenvalue $\tilde{\mu}^2$ with a certain boundary condition for $\phi(\tilde{v})$.
Then, the solution of Eq. (\ref{radial1}) for the radial mode $\Phi(\tilde{u})$ is 
\begin{eqnarray}
\Phi(\tilde{u})\sim A e^{\tilde{u}/\tilde{L}} K_\nu (i\tilde{\mu} \tilde{L}e^{\tilde{u}/\tilde{L}}) + B e^{\tilde{u}/\tilde{L}}I_\nu (i\tilde{\mu} \tilde{L}e^{\tilde{u}/\tilde{L}}) \label{lorentz} 
\end{eqnarray}
with $\nu \equiv \sqrt{1+( m_{\rm r}\tilde{L} )^2}$.
The spectrum of the renormalized mass $m_r$ should be determined for the boundary condition at the UV limit, $\tilde{u}=\log \epsilon$, detail of which will be discussed later.
The competition between two scales $m_{\rm r}$ and $\tilde{\mu}$ generates a nontrivial scaling dimension $\Delta \equiv 1\pm \nu$ for $\Phi$. 
Here, we should remark that the solution (\ref{lorentz}) has the time like signature.

%\subsection{RG trajectory of Wilson NRG and Geodesics}

In the Wilson NRG, the extended translation operation can be viewed as the RG flow of the Wilson Hamiltonian.
Let us consider the relative role of the translations of $\tilde{u}$ and $\tilde{v}$ coordinates.
In Eqs. (\ref{rindler1}) and (\ref{radial1}), the  translation operations for $\tilde{u}$ and $\tilde{v}$ can be performed independently.
In other words, there are two quantum number $\mu$ and $m_{\rm r}$, which can be determined by certain boundary conditions respectively for $\tilde{u}$ and $\tilde{v}$ directions, reflecting that the extra-dimension $\tilde{u}$ is added to the original 1+1D problem.
However, it should be noted that the uniform translation $\tilde{u}\to \tilde{u}+\alpha$ and $\tilde{v}\to \tilde{v}+\alpha$ keeps the metric (\ref{adsrindler}) scale-invariant.
In the level of the equation of motion, thus, this uniform translation keeps both of (\ref{rindler1}) and (\ref{radial1}) invariant,  combined only with  $\tilde{\mu} e^{\alpha/\tilde{L}} \to \tilde{\mu}$.

In order to see the role of the trajectory of the uniform translation, we consider the geodesics for a fixed time ($\eta= $const), i.e. $ds^2= du^2 + e^{2(v-u)}dv^2$.
The nonzero Christoffel symbols are obtained as $\Gamma_{22}^1 = e^{2(\tilde{v}-\tilde{u})/\tilde{L}}/\tilde{L}$, $\Gamma_{12}^2=-1/\tilde{L}$,  $\Gamma_{22}^2=1/\tilde{L}$.
Thus, the geodesics equations are
\begin{eqnarray}
\tilde{L}\frac{d^2 \tilde{u}}{dt^2} + e^{2(\tilde{v}-\tilde{u})/\tilde{L}}(\frac{d \tilde{v}}{dt})^2=0 \\
\tilde{L}\frac{d^2 \tilde{v}}{dt^2} -2 \frac{d \tilde{v}}{dt}\frac{d \tilde{u}}{dt} +(\frac{d \tilde{v}}{dt})^2=0
\end{eqnarray}
A formal solution is 
\begin{equation}
\tilde{v}=-\tilde{L}\log|\alpha + t|+\beta, \quad \tilde{u}=-\tilde{L}\log|\alpha + t|+\beta -i\pi/2
\label{trajectory}
\end{equation}
where $\alpha$ and $\beta$ are integral constants.
This solution satisfies $\tilde{u}=\tilde{v}- i\pi \tilde{L}/2$($e^{2\tilde{u}/\tilde{L}}= -e^{2\tilde{v}/\tilde{L}}$), which corresponds to the uniform translation operation.
Then, Eq. (\ref{radial1}) along the geodesics  $\tilde{u}=\tilde{v}- i\pi \tilde{L}/2$ becomes 
\begin{eqnarray}
(\partial_{\tilde{v}}^2 + \frac{2}{\tilde{L}}\partial_{\tilde{v}}-m_{\rm r}^2-\tilde{\mu}^2e^{2\tilde{v}/\tilde{L}})\Phi(\tilde{v})=0 \label{radial2}
\end{eqnarray}
which reproduces the usual field equation of the radial mode of the euclidean signature.
In Eq. (\ref{radial2}), it is easily checked that $\tilde{\mu}$  exhibits the same exponential dependence as Eq. (\ref{rindler1}).
Along the geodesics (\ref{trajectory}), the boundary condition for $\Phi(\tilde{v})$  and  the renormalized mass $m_{\rm r}$ should be constructed so as to reproduce the proper spectrum of $\tilde{\mu}$ determined by $\phi(\tilde{v})$.
This may suggest that the effect of impurity spin within the single particle state can be renormalized into the boundary condition of $\Phi$ and its effective mass $m_{\rm r}$.

Here, it should be commented on  the holographic entanglement entropy calculated by the physical geodesics at $\eta=0$\cite{ee}, which is given by
\begin{equation}
z^2+x^2=\ell^2 , \label{geod1}
\end{equation}
where $z=Le^{\tilde{u}/\tilde{L}}$ and $x=Le^{\tilde{v}/\tilde{L}}$.
The boundary theory(in the sense of AdS/CFT) is located at $z=0$ and $\ell$ is the length of the system part. 
By measuring the length of along the geodesic (\ref{geod1}),  the entanglement entropy is estimated as $S\sim \frac{c}{6}\log\frac{\ell}{\varepsilon} + g $\cite{affleck2}, where $g$ is a boundary entropy related to the tension at the boundary\cite{HBCFT}.
By contrast, the formal solution (\ref{trajectory}) may be ill defined in the sense of the gravity. 
However,  it might be possible to interpret its physical meaning as follows. 
We can take the system part  as the infinitesimal region of the cutoff scale surrounding $x=0$;
we consider $\ell \to \varepsilon$ of the physical geodesics (\ref{geod1}).
Then, the boundary condition for both of $\phi(\tilde{v})$ and $\Phi(\tilde{v})$ is set up at the geodesics (\ref{geod1}) of the infinitesimal $\ell$.
In the bulk bath region(free electron part) rather than the system part(impurity site), we have  $z^2=\varepsilon^2-x^2 \simeq -x^2$ can be viewed as the Wilson NRG trajectory (\ref{trajectory}) relating $\phi(\tilde{v})$ and $\Phi(\tilde{u})$.
It might be expected that the effect of the impurity spin would be absorbed into the renormalized mass $m_{\rm r}$, if we look at the impurity site sufficiently away from $x=z=0$.
Of course, a more precise analysis is necessary for the thorough understanding.

We discuss the orthonomral basis of (\ref{rindler1}) in terms of the Wilson NRG.
As was mentioned, the spectrum of $\tilde{\mu}$ can be obtained by the translation operation of $\tilde{v}$.
In addition, the inner product of the mode functions is defined by (\ref{norm}),  which is different from the usual field theory where the inner product is  defined as the integral on the hypersurface of time $\eta=const$.
Thus,¡¡we need another treatment of $\{\phi\}$ from the usual field theory based on the canonical quantization. 

We check the orthogonality of a set of wavefunctions $\{\phi\}$ that is generated by the translation operation in the $\tilde{v}$ direction. 
The inner product (\ref{norm}) of $\phi(\xi-\tilde{v})$ and $\phi(\xi-\tilde{v}')$ can be explicitly calculated as
\begin{eqnarray}
 \int_{-\infty}^\infty \frac{d\xi}{\tilde{L}}  e^{2\xi/\tilde{L}} K_{-ik_F\tilde{L}}(\tilde{\mu} \tilde{L}e^{(\xi-v')/\tilde{L}})K_{ik_F\tilde{L}}(\tilde{\mu} \tilde{L}e^{(\xi-v)/\tilde{L}}) \nonumber \\
= \frac{-i\pi e^{i k_F(v+v')}}{2\tilde{L}^2\tilde{\mu}^2\sinh \pi k_F\tilde{L}} \frac{e^{-2ik_Fv'}-b^{-2ik_Fv}}{e^{-2v'/\tilde{L}}-e^{-2v/\tilde{L}}}
\end{eqnarray}
This overlap integral does not orthogonal for arbitrary $v-v'$. 
However, if the continuous Rindler space is discretized into a lattice,
\begin{equation}
\tilde{v}-\tilde{v}'=\frac{\pi}{k_F}n, \qquad n = \pm 1,\pm2, \cdots,
\end{equation}
the wavefunctions on this lattice point become orthogonal. 
Taking the limit $v'\to v$, we can determine the normalization constant.
Then, setting up $v=0$ as the origin of the lattice,  we obtain the orthonormal basis with respect to the lattice translation as 
\begin{equation}
\phi_0(\tilde{v}-an ) =  \sqrt{\frac{2\sinh \pi k_F\tilde{L}}{\pi k_F \tilde{L}} } \tilde{\mu}_0\tilde{L} e^{-\frac{an}{\tilde{L}}}K_{ik_F\tilde{L}}(\tilde{\mu}_0 \tilde{L} e^{\frac{-an}{\tilde{L}}} e^{\frac{\tilde{v}}{\tilde{L}}})
\end{equation}
with the microscopic lattice space  $a\equiv \frac{\pi}{k_F}$.
As was pointed out before,  the mass eigenvalue $\tilde{\mu}$ shows the exponential dependence.
We can thus read 
\begin{equation}
\tilde{\mu}_n\equiv \tilde{\mu}_0\exp(-an/\tilde{L})\qquad n\in Z_n
\end{equation}
where we have assigned $\mu_0$ for a certain mass eigenvalue of order of unity corresponding to $\tilde{v}=0$.
Therefore, the lattice formulation is essential in the Wilson NRG, although the logarithmic discretization is technically  introduced for the purpose of numerical computation.
The arbitrary field $\hat{\phi}$ on the $\tilde{v}$ lattice point can be expanded as 
$\hat{\phi}(\tilde{v})= \sum_n \phi_0(\tilde{v}-an)a_n$, 
where $a_n$ is the annihilation operator of a particle of the $n$th mass mode.
An important point is that  $\tilde{v}$ and mass index of $n$ are involved  on equal footing, suggesting a self-dual structure between the real space and the mass spectrum.

We recover  the basis of the pre-scaled variables from  the ``tilde" variables.
Recall the $\tilde{v}=\frac{\omega}{k_F}v$, $\tilde{u}=\frac{\omega}{k_F}$, $\tilde{\mu}=\frac{k_F}{\omega}\mu$, $\tilde{L}=\frac{\omega}{k_F} L$, where $\tilde{L}$ depends on the frequency $\omega$ through $\tilde{L}\propto \omega L$.
Thus, if $\tilde{L}$ is fixed as in Wilson NRG, the AdS radius $L$ of the original coordinate becomes the frequency dependent.
Indeed, we have
\begin{equation}
\phi_0(v-\frac{\pi}{\omega}n ) =  \sqrt{\frac{\sinh \pi \omega L}{\pi \omega L}} \mu_0L e^{\frac{\pi n}{\omega L}} K_{i\omega L}( \mu_0 L e^{\frac{\pi n }{\omega L}} e^{\frac{v}{L}}), 
\label{prescaledfield}
\end{equation}
which just reflects that the independent parameters of the equation of motion  are $\omega L $ and $\mu_0 L$ in the original AdS$_3$ coordinate.
In addition, recall that the energy scale with respect to $\mu$ can be controlled by the lattice translation. 
Thus, the controlling parameter of the Wilson theory eventually reduces to the $\omega$-dependent AdS radius, which can be related to the deformation parameter $\lambda=2k_F/\omega L$ in the lattice Hamiltonian (\ref{Hamiltonian}).
As was discussed previously, the Wilson NRG basically treats the spectrum of $\mu$ for the fixed $\omega L$, while the conventional Rindler theory deals with the complete set of the fields with respect to $\omega$ for $\mu$.
In order to cast the wavefunction of the Wilson NRG into the Rindler field, we impose $e^{\frac{\pi}{\omega L}n}$ on  $L$ in Eq. (\ref{prescaledfield}), i.e.  we consider  scale transformation $L'=e^{\frac{\pi}{\omega L}n} L$ combined with $\omega'=e^{-\frac{\pi}{\omega L}n} \omega$ and $v'= e^{-\frac{\pi}{\omega L}n} v$.
This transformation yields the Rindler field (\ref{prescaledfield}) having the spectrum of $\omega'$ for a fixed mass scale $\mu_0$.
Instead, the AdS radius $L'$ also becomes $\omega$-dependent.
It is therefore concluded that  the orthonormal basis in the Wilson NRG are associated with the boundary Rindler field of the mode-dependent AdS Radius $L$.

We finally discuss the role of the radial field $\Phi(\tilde{v})$.
The scalability of the $\tilde{\mu}^2 e^{2\tilde{v}/\tilde{L}}$ term in Eq. (\ref{radial2}) is the same as $\phi(\tilde{v})$.
Thus we can reproduce the spectrum of $\tilde{\mu}$ with the lattice space of $a$, once the UV limit of $\Phi(\tilde{v})$ is determined in consistency with $\phi(\tilde{v})$.
Then, the solution of (\ref{radial2}) in $\tilde{v}$ coordinate is given by
\begin{eqnarray}
\Phi(\tilde{v})&\sim& A e^{\tilde{v}/\tilde{L}} K_\nu (\tilde{\mu} \tilde{L}e^{\tilde{v}/\tilde{L}}) +  B e^{\tilde{v}/\tilde{L}}I_\nu (\tilde{\mu} \tilde{L}e^{\tilde{v}/\tilde{L}}) 
\label{euclidradial}
\end{eqnarray}
with $\nu \equiv \sqrt{1+( m_{\rm r}\tilde{L} )^2}$.
Here it should be remarked that Eq. (\ref{euclidradial}) has the euclidean signature.
The scaling dimension of $\Phi(\tilde{v})$  may take a nontrivial value depending on the renormalized mass $m_{\rm r}$, which would be determined by the impurity scattering at the UV limit.
In $v\to \infty$, the modified Bessel function of $I_\nu$ is diversive and thus we can basically assume $B=0$.

Here, we would like to remark that, in this paper, we have not assumed the AdS/CFT explicitly.
We have basically considered the single particle state of the Wilson's effective Hamiltonian (\ref{Hamiltonian}) representing the bulk electrons, and the AdS$_3$ space time was introduced  as a background metric, where the radial mode $\Phi$ is the auxiliary field to control the effective mass scale $\mu$. 
At the present stage, thus, it is subtle to judge whether the Wilson NRG is a  representation of AdS/CFT or not, although we have found the similar structure between them.

In the Kondo problem, nevertheless, the interaction exists only at the UV limit and the IR behavior is described by the local Fermi liquid\cite{nozieres}, which can be correctly extracted by the Wilson NRG computation\cite{Wilson}.
It is well-known that the Kondo problem can be described by the boundary CFT\cite{affleck}, which is also  discussed from the holographic entanglement entropy\cite{boundary}.
Recently, the boundary MERA, which can successfully capture the scaling dimension of the boundary excitation,  leads the similar effective Hamiltonian with the exponential deformation for the uniformly interacting system\cite{boundaryMERA,boundaryMERA2,okub}.
The present approach may provide an interesting view point to illustrate the relationship between the RG description in condensed matter physics and AdS/CFT.

I would like thank Kazutoshi Ohta for useful discussions and comments.
The work is supported by JSPS Grants-in-Aid for Scientific Researches(No. 23540442 and No. 23340190).


\begin{thebibliography}{99}
\bibitem{sign} Sign of $n$ is inverted from that in Ref. \cite{ON} for the later convenience.
\bibitem{boundary} Strictly speaking, the translational invariance is broken at the UV boundary as in Ref. \cite{ON}.
This scattering at the UV boundary causes a phase shift of the wavefunction, which may be a possible source of the renomalization of the mass term $m_{\rm r}$.
\end{thebibliography}
\end{document}